\documentclass[twocolumn,epjc3]{svjour3}  
\smartqed  
\RequirePackage{graphicx}
\RequirePackage[T1]{fontenc}
\usepackage{amssymb}
\usepackage{amsmath}
\usepackage{hyperref}

\begin{document}

\title{Resolution of curvature singularities from quantum mechanical and loop perspective}

\titlerunning{Resolution of curvature singularities}        

\author{T. Tahamtan\thanksref{e1,addr1}
        \and
        O. Sv\'{\i}tek\thanksref{e2,addr2} 
}

\thankstext{e1}{e-mail: tayabeh.tahamtan@emu.edu.tr}
\thankstext{e2}{e-mail: ota@matfyz.cz}

\institute{Department of Physics, Eastern Mediterranean University, G. Magusa,
north Cyprus, Mersin 10, Turkey \label{addr1}
           \and
           Institute of Theoretical Physics, Faculty of Mathematics and Physics,
Charles University in Prague, V Hole\v{s}ovi\v{c}k\'{a}ch 2, 180 00 Prague
8, Czech Republic \label{addr2}
}

\date{Received: date / Accepted: date}

\maketitle

\begin{abstract}
We analyze the persistence of curvature singularities when analyzed using quantum theory. First, quantum test particles obeying the Klein-Gordon and Chandrasekhar-Dirac equation are used to probe the classical timelike naked singularity. We show that the classical singularity is felt even by our quantum probes. Next, we use loop quantization to resolve singularity hidden beneath the horizon. The singularity is resolved in this case.
\end{abstract}

\section{Introduction}

One of the important predictions of the Einstein's theory of general
relativity is the formation of spacetime singularities. In classical general
relativity, singularities are defined as points in which the evolution of
timelike or null geodesics is not defined after a finite proper time. According to
the classification of the classical singularities devised by Ellis and
Schmidt \cite{01}, scalar curvature singularities are the most strongest one in the
sense that the spacetime posses incomplete geodesics ending in them and all the physical quantities
such as the gravitational field (scalars formed from curvature tensor), energy density and tidal forces diverge at
the singular point. 

But such divergence of physical quantities signify the breakdown of predictive power of classical general relativity. If these singularities are covered by horizon (as supposed by Cosmic Censorship Conjecture) then at least the physically most relevant region of spacetime is under control. Naked singularities (those not covered by horizon), on the other hand, provide an observer with causal access to the region of diverging quantities and should be avoided. However, even singularities covered by the horizon can be accessed by an infalling observer and, more importantly, we would like to have a theory that lacks divergences, at least effectively.

The natural direction for resolving the problem of singularities in classical theory is investigating their persistence in quantum picture. Although we do not have a final quantum theory of gravity we still have several tools for analyzing quantum singularities. The first approach relies on examining properties of quantum particle wave functions on the background represented by the studied geometry. This is a frequently used technique based on well understood properties of operators on a Hilbert space. To move further, one might proceed to using quantum fields and possibly even the backreaction of background geometry using semiclassical Einstein equations with suitably regularized stress energy tensor. Finally, one can apply quantization of the geometry itself. The last approach is in principle the most precise but relies on the selected quantization method and we have no generally accepted one in case of gravity.

Quantum singularities were studied for different specific situations (and using also generalizations), mainly using the first approach \cite{6,7,8,9,10,11,12,13,140,14,15,16,17}. Recently, singularities in f(R) gravity were investigated in the presence of linear electromagnetic field \cite{170}.

We will apply two of the above mentioned approaches for analysis of singularity in case of the general metric of global monopole \cite{1}, which is determined by two parameters - one characterizing the "Schwarzschild-type mass" and the other one the deficit of solid angle. The singularity is generally covered by single horizon but the class of metrics also contains, as a special case, a naked singularity which is analyzed from quantum mechanical point of view using the technique of Horowitz and Marolf \cite{4} (who continued the pioneering work of Wald \cite{3}). This method for analyzing timelike singularities is based on investigation of self-adjoint extensions of the evolution operator associated with the given wave equation. If it is unique the spacetime is deemed quantum mechanically non-singular. The analysis is carried out for relativistic quantum particle wave equations on a fixed background. Specifically, we review the previous results for Klein-Gordon equation and show the calculation using Newman-Penrose formalism for the Dirac equation, both in the case of pure global monopole with naked singularity for which the method was developed.

But as already mentioned, the most reliable method when trying to investigate the possible removal of the singularities from geometry is quantum gravity. Here we have selected loop quantization method inspired by \cite{22,23,24}, where the spacetime beneath the horizon (in the non-naked subclass) is isometric to the Kantowski-Sachs cosmology. Then one can apply methods from Loop Quantum Cosmology (LQC), that are based on loop quantization on the restricted configuration space. In this way, the results for resolution of initial cosmological singularity are translated to statements about the singularity at the origin $r=0$.

\section{The General Metric for Global monopole}

It is well known that different types of non-standard topological objects may have been
formed during initial Universe evolution, such as domain walls, cosmic strings and
monopoles \cite{1,2}. The basic idea is that these topological defects
have formed as a result of a breakdown of local or global gauge symmetries. The simplest
model that gives rise to global monopole is described by the Lagrangian

\begin{equation}
L=\frac{1}{2}\partial _{\mu }\phi ^{a}\partial ^{\mu }\phi ^{a},
\end{equation}%
where $\phi ^{a}$ is a triplet of scalar fields, $a=1,2,3.$ The model has a
global $O\left( 3\right) $ symmetry, which is spontaneously broken to $%
U\left( 1\right) $. The field configuration describing the monopole is

\begin{equation*}
\phi ^{a}=\eta \frac{x^{a}}{r}
\end{equation*}%
where $x^{a}x^{a}=r^{2}$. We assume that underlying geometry is general
static spherically symmetric described by the line element 
\begin{equation}\label{general}
ds^{2}=-B\left( r\right) dt^{2}+\frac{dr^{2}}{A\left( r\right) }+r^{2}\left(
d\theta ^{2}+\sin ^{2}\theta d\phi ^{2}\right) ,
\end{equation}%
with the usual relation between the spherical coordinates, $r,\theta ,\phi $
and the Cartesian coordinates $x^{a}$. The Lagrangian for the above given
field configuration simplifies in the following way%
\begin{equation}\label{scalar_L}
L=\frac{1}{2}\left( \partial _{\theta }\phi ^{a}\partial ^{\theta }\phi
^{a}+\partial _{\phi }\phi ^{a}\partial ^{\phi }\phi ^{a}\right) =\frac{\eta
^{2}}{r^{2}},
\end{equation}%
and the diagonal energy momentum tensor is given by these components%
\begin{equation}
T_{t}^{t}=T_{r}^{r}=-\frac{\eta ^{2}}{r^{2}},\text{ \ \ }T_{\theta }^{\theta
}=T_{\phi }^{\phi }=0.\text{\ }
\end{equation}%
The general solution of the Einstein equations with this $T_{\mu }^{\nu }$ is

\begin{equation}
B=A^{-1}=1-8\pi G\eta ^{2}-\frac{2GM}{r}
\end{equation}%
where $M$ is a constant of integration. The metric describes a black hole of
mass $M$, carrying a global monopole charge characterized by $\eta$. Such a black hole can be formed
if a global monopole is swallowed by an ordinary black hole \cite{1}.

The Kretschmann scalar which indicates the formation of curvature
singularity is given by%
\begin{equation}
\mathcal{K}=\frac{48M^{2}G^{2}}{r^{6}}+\frac{128M\pi G^{2}\eta ^{2}}{r^{5}}+%
\frac{256\pi ^{2}G^{2}\eta ^{4}}{r^{4}}.
\end{equation}%
It is obvious that $r=0$ is a typical central curvature singularity (scalar curvature singularity according to above mentioned classification) and the dominant contribution comes from term corresponding to black hole mass $M$. If $M > 0$ the singularity is evidently spacelike and covered by a single horizon.

\section{Global monopole and its singularity}

If we assume that the mass term is negligible on the astrophysical scale or vanishing, we
will have%
\begin{equation}\label{naked}
ds^{2}=-\left( 1-8\pi G\eta ^{2}\right) dt^{2}+\frac{dr^{2}}{\left( 1-8\pi
G\eta ^{2}\right) }+r^{2}d\Omega^{2} ,
\end{equation}%
For simplicity we choose $\alpha ^{2}=1-8\pi G\eta ^{2}$ and by rescaling $r$
and $t$ variables, we can rewrite the monopole metric as 
\begin{equation}\label{monopole}
ds^{2}=-dt^{2}+dr^{2}+\alpha ^{2}r^{2}\left( d\theta ^{2}+\sin ^{2}\theta
d\varphi ^{2}\right) ,
\end{equation}%
If we calculate the Kretschmann scalar,%
\begin{equation*}
\mathcal{K}=4\frac{\left( \alpha ^{2}-1\right) ^{2}}{r^{4}\alpha ^{4}}.
\end{equation*}%
still there is a weaker singularity at $r=0$. From the metric (\ref{naked}) one can immediately see that the singularity is timelike. This time, because our simplified metric does not have the horizon the singularity is naked.

\section{Naked Singularity}

As mentioned in the Introduction naked singularity poses a serious problems and its resolution would be desirable. In this section, the occurrence of naked singularities in global monopole
will be analyzed from quantum mechanical point of view. In probing the
singularity, quantum test particles obeying the Klein-Gordon and
Dirac equations are used. The reason for using two different types of fields is
to clarify whether the classical singularity is sensitive to spin of the
fields.

According to Horowitz and Marolf (HM) \cite{4}, the singular character of the spacetime is defined as the ambiguity in
the evolution of the wave functions. That is to say, the singular character
is determined based on the number of self-adjoint extensions of the
evolution operator to the entire Hilbert space. If the extension is unique, it is said
that the spacetime is quantum mechanically regular. The brief review of the method follows:

Consider a static spacetime $\left( \mathcal{M},g_{\mu \nu }\right) $\ with
a timelike Killing vector field $\xi ^{\mu }$. Let $t$ denote the Killing
parameter and $\Sigma $\ denote a static slice. The Klein-Gordon equation in
this space is

\begin{equation}
\left( \nabla ^{\mu }\nabla _{\mu }-M^{2}\right) \psi =0.
\end{equation}%
This equation can be written in the form

\begin{equation}
\frac{\partial ^{2}\psi }{\partial t^{2}}=\sqrt{f}D^{i}\left( \sqrt{f}%
D_{i}\psi \right) -fM^{2}\psi =-A\psi ,
\end{equation}%
in which $f=-\xi ^{\mu }\xi _{\mu }$ and $D_{i}$ is the spatial covariant
derivative on $\Sigma $. We assume that the Hilbert space $\mathcal{H}=
L^{2}\left( \Sigma, \mu \right)$ is the space of square integrable
functions on $\Sigma $ with appropriate measure $\mu $. Initially the operator $A$ is defined on smooth functions with compact support $C_{0}^{\infty}(\Sigma)$. Since the operator $A$ is real,
positive and symmetric its self-adjoint extensions always exist.
If it has a unique extension $A_{E},$ then $A$ is called essentially
self-adjoint \cite{18,19,20}. Accordingly, the Klein-Gordon equation for
a free particle satisfies

\begin{equation}
i\frac{d\psi}{dt}=\sqrt{A_{E}}\psi,
\end{equation}
with the solution

\begin{equation}
\psi \left( t\right) =\exp \left[ -it\sqrt{A_{E}}\right] \psi \left(
0\right) .
\end{equation}%
If $A$ is not essentially self-adjoint, the future time evolution of the
wave function (12) is ambiguous. Then, HM criterion defines the spacetime
as quantum mechanically singular. However, if there is only a single
self-adjoint extension, the operator $A$ is said to be essentially
self-adjoint and the quantum evolution described by equation (12) is
uniquely determined by the initial conditions. According to the HM
criterion, this spacetime is said to be quantum mechanically non-singular.
In order to determine the number of self-adjoint extensions, the concept of
deficiency indices is used. The deficiency subspaces $N_{\pm }$ are defined
by ( see Ref.\cite{5}  for a detailed mathematical background),

\begin{align}
N_{+}& =\{\psi \in D(A^{\ast }),\text{\ \ }A^{\ast }\psi =Z_{+}\psi
,\text{\ \ }ImZ_{+}>0\}\notag\\
&\text{\ \ with dimension }n_{+} \\
N_{-}& =\{\psi \in D(A^{\ast }),\text{ \ \ }A^{\ast }\psi =Z_{-}\psi
,\text{\ \ }ImZ_{-}<0\}\notag\\
&\text{\ \ with dimension }n_{-}  \notag
\end{align}%
The dimensions $\left( \text{ }n_{+},n_{-}\right) $ are the deficiency
indices of the operator $A$. The indices $n_{+}(n_{-})$ are completely
independent of the choice of $Z_{+}(Z_{-})$ depending only on whether $Z$
lies in the upper (lower) half complex plane. Generally one takes $%
Z_{+}=i\lambda $ and $Z_{-}=-i\lambda $ , where $\lambda $ is an arbitrary
positive constant necessary for dimensional reasons. The determination of
deficiency indices then reduces to counting the number of solutions of $%
A^{\ast }\psi =Z\psi $ ; (for $\lambda =1$),

\begin{equation}\label{adjointness}
A^{\ast }\psi \pm i\psi =0
\end{equation}%
that belong to the Hilbert space $\mathcal{H}$. If there are no square
integrable solutions ( i.e. $n_{+}=n_{-}=0)$, the operator $A$ possesses a
unique self-adjoint extension and it is essentially self-adjoint.
Consequently, a sufficient condition for the operator $A$ to be essentially
self-adjoint is to find only solutions satisfying Eq. (14) that do not
belong to the Hilbert space.

\subsection{Klein - Gordon Fields}

The Klein-Gordon equation for a massless scalar particle is given by,

\begin{equation}
\square \psi =g^{-1/2}\partial _{\mu }\left[ g^{1/2}g^{\mu \nu }\partial
_{\nu }\right] \psi =M^{2}\psi .
\end{equation}%
For the metric (\ref{monopole}), the Klein-Gordon equation becomes,

\begin{equation}
\frac{\partial ^{2}\psi }{\partial t^{2}}=-\left\{ \frac{\partial ^{2}\psi }{%
\partial r^{2}}+\frac{1}{r^{2}\alpha ^{2}}\frac{\partial ^{2}\psi }{\partial
\theta ^{2}}+\frac{1}{r^{2}\alpha ^{2}\sin ^{2}\theta }\frac{\partial
^{2}\psi }{\partial \varphi ^{2}}+\right.
\end{equation}
\[
\left. \qquad\qquad +\frac{\cos \theta }{r^{2}\alpha ^{2}\sin
\theta }\frac{\partial \psi }{\partial \theta }+\frac{2}{r}\frac{\partial
\psi }{\partial r}\right\} .
\]
In analogy with the equation (10), the spatial operator $A$ is

\begin{equation}
\emph{A}=\left\{ \frac{\partial ^{2}}{\partial r^{2}}+\frac{1}{r^{2}\alpha
^{2}}\frac{\partial ^{2}}{\partial \theta ^{2}}+\frac{1}{r^{2}\alpha
^{2}\sin ^{2}\theta }\frac{\partial ^{2}}{\partial \varphi ^{2}}\right. +
\end{equation}
\[
\left. \qquad\qquad+\frac{\cos
\theta }{r^{2}\alpha ^{2}\sin \theta }\frac{\partial }{\partial \theta }+%
\frac{2}{r}\frac{\partial }{\partial r}\right\} .
\]
and the equation to be solved is $\left( \emph{A}^{\ast }\pm i\right) \psi
=0$. Using separation of variables, $\psi =R\left( r\right) Y_{l}^{m}\left(
\theta ,\varphi \right) $, we get the radial portion of equation (14) as,

\begin{equation}
\frac{d^{2}R\left( r\right) }{dr^{2}}+\frac{2}{r}\frac{dR\left( r\right) }{dr%
}+\left( \frac{-l\left( l+1\right) }{r^{2}\alpha ^{2}}\pm i\right) R\left(
r\right) =0.
\end{equation}

The square integrability of the above solution is checked by calculating the
squared norm of the above solution in which the function space on each $t=$
constant hypersurface $\Sigma $ is defined as $\mathcal{H=}L^{2}\left(
\Sigma ,\mu \right) $ where $\mu $\ is the measure given by the spatial metric
volume element. 

We easily recover the results showed in \cite{12}: The spacetime of global
monopole remains singular in the view of relativistic quantum mechanics: the future of a
given initial wave packet obeying the Klein-Gordon equation is not generally well determined, similarly
to the future of a classical particle which reaches the classical singularity at $r=0$.

\subsection{Dirac Fields}

The Newman-Penrose formalism will be used here to analyze massless
Dirac particle propagating in the space of global monopole. The
signature of the metric (\ref{monopole}) is changed to $-2$ in order to use the Dirac
 equation in Newman-Penrose formalism. Thus, the metric is given by,%
\begin{equation}\label{monopole-dirac}
ds^{2}=dt^{2}-dr^{2}-r^{2}\alpha ^{2}\left( d\theta ^{2}+\sin ^{2}\theta
d\varphi ^{2}\right) .
\end{equation}%
The Chandrasekhar-Dirac (CD) \cite{140} equations in Newman-Penrose formalism are given
by

\begin{eqnarray}
\left( D+\epsilon -\rho \right) F_{1}+\left( \bar{\delta}+\pi -\alpha
\right) F_{2} &=&0, \\
\left( \nabla +\mu -\gamma \right) F_{2}+\left( \delta +\beta -\tau \right)
F_{1} &=&0,  \notag \\
\left( D+\bar{\epsilon}-\bar{\rho}\right) G_{2}-\left( \delta +\bar{\pi}-%
\bar{\alpha}\right) G_{1} &=&0,  \notag \\
\left( \nabla +\bar{\mu}-\bar{\gamma}\right) G_{1}-\left( \bar{\delta}+\bar{%
\beta}-\bar{\tau}\right) G_{2} &=&0,  \notag
\end{eqnarray}%
where $F_{1},F_{2},G_{1}$ and $G_{2}$ are the components of the wave
function, $\epsilon ,\rho ,\pi ,\alpha ,\mu ,\gamma ,\beta $ and $\tau $ are
the spin coefficients to be found and the "bar" denotes complex conjugation.
The null tetrad vectors for the metric (\ref{monopole-dirac}) are defined by%
\begin{eqnarray}
l^{a} &=&\left( 1,1,0,0\right) , \\
n^{a} &=&\left( \frac{1}{2},-\frac{1}{2},0,0\right) ,  \notag \\
m^{a} &=&\frac{1}{\sqrt{2}}\left( 0,0,\frac{1}{\alpha r},\frac{i}{r\alpha
\sin \theta }\right) .  \notag
\end{eqnarray}%
The directional derivatives in the Dirac equation are defined by $%
D=l^{a}\partial _{a},\nabla =n^{a}\partial _{a}$ and $\delta =m^{a}\partial
_{a}.$ We define operators in the following way

\begin{eqnarray}
\mathbf{D}_{0} &=&D  \notag \\
\mathbf{D}_{0}^{\dagger } &=&-2\nabla  \\
\mathbf{L}_{0}^{\dagger } &=&\sqrt{2}r\text{ }\alpha \delta \text{ and }%
\mathbf{L}_{1}^{\dagger }=\mathbf{L}_{0}^{\dagger }+\frac{\cot \theta }{2} 
\notag \\
\mathbf{L}_{0} &=&\sqrt{2}r\alpha \text{ }\bar{\delta}\text{ and }\mathbf{L}%
_{1}=\mathbf{L}_{0}+\frac{\cot \theta }{2}  \notag
\end{eqnarray}%
The nonzero spin coefficients are,

\begin{equation}
\mu =-\frac{1}{2r},\rho =-\frac{1}{r},\beta =-\mathbf{\alpha }=\frac{1}{2%
\sqrt{2}}\frac{\cot \theta }{r\alpha }.
\end{equation}%
Substituting nonzero spin coefficients and the definitions of the operators
given above into the CD equations leads to

\begin{gather}
\left( \mathbf{D}_{0}+\frac{1}{r}\right) F_{1}+\frac{1}{r\alpha \sqrt{2}}%
\mathbf{L}_{1}F_{2}=0,  \notag \\
-\frac{1}{2}\left( \mathbf{D}_{0}^{\dagger }+\frac{1}{r}\right) F_{2}+\frac{1%
}{r\alpha \sqrt{2}}\mathbf{L}_{1}^{\dagger }F_{1}=0,  \notag \\
\left( \mathbf{D}_{0}+\frac{1}{r}\right) G_{2}-\frac{1}{r\alpha \sqrt{2}}%
\mathbf{L}_{1}^{\dagger }G_{1}=0,  \notag \\
\frac{1}{2}\left( \mathbf{D}_{0}^{\dagger }+\frac{1}{r}\right) G_{1}+\frac{1%
}{r\alpha \sqrt{2}}\mathbf{L}_{1}G_{2}=0.\label{NP-Dirac}
\end{gather}%
For the solution of the CD equations, we assume separable solution in the
form of%
\begin{eqnarray}
F_{1} &=&f_{1}(r)Y_{1}(\theta )e^{i\left( kt+m\varphi \right) },  \notag\\
F_{2} &=&f_{2}(r)Y_{2}(\theta )e^{i\left( kt+m\varphi \right) },  \notag \\
G_{1} &=&g_{1}(r)Y_{3}(\theta )e^{i\left( kt+m\varphi \right) },  \notag \\
G_{2} &=&g_{2}(r)Y_{4}(\theta )e^{i\left( kt+m\varphi \right) }.  \label{separation}
\end{eqnarray}%
Here $\left\{ f_{1},f_{2},g_{1},g_{2}\right\} $ and $\left\{
Y_{1},Y_{2},Y_{3},Y_{4}\right\} $ are functions of $r$ and $\theta $
respectively, $m$ is the azimuthal quantum number and $k$ is the frequency of the
Dirac spinor, which is assumed to be positive and real. By substituting (\ref{separation}) in (\ref{NP-Dirac}) we will see that with these
assumptions%
\begin{eqnarray}
\text{\ }f_{1}(r) &=&g_{2}(r)\text{ \ \ \ \ and \ \ \ }f_{2}(r)=g_{1}(r)%
\text{\ \ }, \\
Y_{1}(\theta ) &=&Y_{3}(\theta )\text{ \ \ \ \ and \ \ \ }Y_{2}(\theta
)=Y_{4}(\theta )
\end{eqnarray}%
Dirac equation reduces to two equations. The radial part of the Dirac equations become

\begin{gather}
\left( \mathbf{D}_{0}+\frac{1}{r}\right) f_{1}\left( r\right) =\frac{\lambda 
}{r\alpha \sqrt{2}}f_{2}\left( r\right) , \label{reduced}\\
\frac{1}{2}\left( \mathbf{D}_{0}^{\dagger }+\frac{1}{r}\right) f_{2}\left(
r\right) =\frac{\lambda }{r\alpha \sqrt{2}}f_{1}\left( r\right) .  \notag
\end{gather}%
where $\lambda $\ comes from separation of variables. We further assume that

\begin{eqnarray*}
f_{1}\left( r\right)  &=&\frac{\Psi _{1}\left( r\right) }{r}, \\
f_{2}\left( r\right)  &=&\frac{\sqrt{2}\Psi _{2}\left( r\right) }{r},
\end{eqnarray*}%
then equation (\ref{reduced}) transforms into,

\begin{gather}
\mathbf{D}_{0}\Psi _{1}=\frac{\lambda ^{^{\prime }}}{r}\Psi _{2}, \\
\mathbf{D}_{0}^{\dagger }\Psi _{2}=\frac{\lambda ^{^{\prime }}}{r}\Psi _{1}.
\notag
\end{gather}%
where $\lambda ^{^{\prime }}=\frac{\lambda }{\alpha },$ so we will have%
\begin{eqnarray}
\left( \frac{d}{dr}+ik\right) \Psi _{1} &=&\frac{\lambda ^{^{\prime }}}{r}%
\Psi _{2}, \\
\left( \frac{d}{dr}-ik\right) \Psi _{2} &=&\frac{\lambda ^{^{\prime }}}{r}%
\Psi _{1},  \notag
\end{eqnarray}%
In order to write the above equation in a more compact form we combine the
solutions in the following way,%
\begin{eqnarray*}
Z_{+} &=&\Psi _{1}+\Psi _{2}, \\
Z_{-} &=&\Psi _{2}-\Psi _{1}.
\end{eqnarray*}%
After doing some calculations we end up with a pair of one-dimensional Schr\"{o}dinger-like wave equations with effective potentials,

\begin{gather}
\left( \frac{d^{2}}{dr^{2}}+k^{2}\right) Z_{\pm }=V_{\pm }Z_{\pm }, \\
V_{\pm }=\frac{\lambda ^{^{\prime }2}}{r^{2}}\mp \frac{\lambda ^{^{\prime }}%
}{r^{2}}.
\end{gather}%
In analogy with the equation (10), the spatial operator $A$ for the massless
case is

\begin{equation*}
A=-\frac{d^{2}}{dr^{2}}+V_{\pm },
\end{equation*}%
so we have (see (\ref{adjointness}))

\begin{equation}
\left( \frac{d^{2}}{dr^{2}}-\left[ \frac{\lambda ^{^{\prime }2}}{r^{2}}\mp 
\frac{\lambda ^{^{\prime }}}{r^{2}}\right] \mp i\right) \psi_{\pm }=0.
\end{equation}%
The solutions of the above equations are expressible using Bessel functions
of the first and second kind in the following way%
\begin{eqnarray}
\psi_{+} &=&C_{1}\sqrt{r}J\left( \lambda ^{^{\prime }}-\frac{1}{2},\frac{r}{%
\sqrt{2}}\left( 1-i\right) \right) +\notag\\
& &C_{2}\sqrt{r}Y\left( \lambda ^{^{\prime}}-\frac{1}{2},\frac{r}{\sqrt{2}}\left( 1-i\right) \right) ,  \notag \\
\psi_{-} &=&C_{1}^{^{\prime }}\sqrt{r}J\left( \lambda ^{^{\prime }}+\frac{1}{2},%
\frac{r}{\sqrt{2}}\left( 1+i\right) \right) +\notag\\
& &C_{2}^{^{\prime }}\sqrt{r} Y\left( \lambda ^{^{\prime }}+\frac{1}{2},\frac{r}{\sqrt{2}}\left(
1+i\right) \right) .
\end{eqnarray}%
Using the asymptotic formulas for Bessel functions when $r\rightarrow \infty 
$ ($Y(\kappa ,z)\approx z^{-1/2}\sin (z-\kappa \pi /2-\pi /4)$ and $J(\kappa
,z)\approx z^{-1/2}\cos (z-\kappa \pi /2-\pi /4)$) and noting the complex
argument in both solutions one can find a combination of constants $%
C_{1},C_{2}$\ or $C_{1}^{\prime },C_{2}^{\prime }$ which is square
integrable near infinity. (But, it is also possible to choose the constants differently so that both solutions are not square integrable!).

When $r\rightarrow 0$ the approximate expressions for Bessel functions ($Y(\kappa ,z)\approx z^{-\kappa }$ for $\kappa\neq 0$, $Y(0 ,z)\approx \ln(z/2)$ and $J(\kappa ,z)\approx z^{\kappa }$) imply that for $C_{2}=0$ and $C_{2}^{\prime }=0$ we have square integrable solution near zero. (Here again if we suppose $C_{1}=0$ and $C_{1}^{\prime }=0$, for $\kappa\geq 3/2$, the solutions are not square integrable!. One could restrict an analysis to only certain wave modes and purposely choose the modes to be quantum regular).

But since we have a solution of equations valid on the whole domain (not just asymptotic forms of equations) we can match the behaviour at zero and infinity. Based on the results we can have solution square integrable over the whole domain and therefore our deficiency indices are nonzero. The operator is not essentially self-adjoint and the spacetime is quantum mechanically singular. 

\section{Quantum Gravity}
Now we are going to investigate the singularity of general global monopole using techniques from loop quantization in the manner of \cite{23}. Consider equation (\ref{general}), for $r<\frac{2GM}{1-8\pi G\eta ^{2}}$. This metric
describes spacetime inside the horizon of a black hole. The coordinate $r$
is timelike and the coordinate $t$ is spatial there; for convenience we rename
them as $r\equiv T$ and $t\equiv r$ with $T\in [0,\frac{2GM}{1-8\pi G\eta
^{2}}]$ and $r\in [-\infty ,+\infty ] $ and the metric becomes%
\begin{equation}
ds^{2}=-\left( \alpha ^{2}-\frac{2GM}{T}\right) dr^{2}+\frac{dT^{2}}{\left(
\alpha ^{2}-\frac{2GM}{T}\right) }+T^{2}d\Omega^{2} ,
\end{equation}%
we eliminate the coefficient of $dT^{2}$ by defining a new temporal variable 
$\tau $ via 
\begin{equation}
d\tau =\frac{dT}{\sqrt{\frac{2GM}{T}-\alpha ^{2}}}.
\end{equation}%
Accordingly, the metric becomes 
\begin{equation}
ds^{2}=-d\tau ^{2}+\left( \frac{2GM}{T}-\alpha ^{2}\right)
dr^{2}+T^{2}\left( d\theta ^{2}+\sin ^{2}\theta d\varphi ^{2}\right) .
\end{equation}%
We introduce two functions $a^{2}\left( \tau \right) \equiv \frac{2Gm}{T}%
-\alpha ^{2}$ and $b^{2}\left( \tau \right) \equiv T^{2}\left( \tau \right) $
and redefine $\tau \equiv t$. The metric becomes 
\begin{equation}
ds^{2}=-dt^{2}+a^{2}\left( t\right) dr^{2}+b^{2}\left( t\right) \left(
d\theta ^{2}+\sin ^{2}\theta d\varphi ^{2}\right) ,
\end{equation}%
this metric describes a homogeneous, anisotropic Kanto{-}wski-Sachs cosmological model with spatial section having
topology $\mathbf{R\times S}^{2}$. From this observation comes the motivation to use LQC approach. In our case $a\left( t\right) $ is a function of $b\left( t\right) $.

\subsection{Classical observables}

The corresponding action for gravity minimally coupled with scalar field (described by (\ref{scalar_L})) can
be written in the form%
\[
S=\frac{1}{16\pi G}\int dtd^{3}xNh^{1/2}\left[ K_{ij}K^{ij}-K^{2}+ \right.
\]
\begin{equation}
\left.\qquad\qquad\qquad +^{\left(3\right) }R-\frac{16\pi G\eta ^{2}}{b^{2}}\right] ,
\end{equation}
by considering the metric (38), the action becomes%
\begin{eqnarray}
S &=&\frac{-1}{8\pi G}\int dt\int_{0}^{R}dr\int_{0}^{2\pi }d\phi \int_{0}^{\pi
}d\theta \sin \theta ab^{2}\times\notag\\
&&\times\left[ \frac{\dot{b}^{2}}{b^{2}}+\frac{2\dot{a}%
\dot{b}}{ab}-\frac{\alpha ^{2}}{b^{2}}\right] .
\end{eqnarray}%
By using the relation between $a$ and $b$, we will be able to write the
action in terms of a single function%
\begin{eqnarray}
S&=&\frac{R\alpha ^{2}}{2G}\int dt\sqrt{\frac{b}{2GM}}\left( 1-\frac{\alpha
^{2}b}{2GM}\right) ^{-1/2}\times\notag\\
&&\times\left[ \dot{b}^{2}+\frac{2GM}{b}\left( 1-\frac{%
\alpha ^{2}b}{2GM}\right) \right] .
\end{eqnarray}%
Now, we will compute the Hamiltonian (Hamiltonian constraint). The
momentum associated to the chosen configuration variable is%
\begin{equation}
p_{b} =\frac{R\alpha ^{2}\dot{b}}{G}\sqrt{\frac{b}{2GM}}\left( 1-\frac{\alpha ^{2}b}{2GM}\right) ^{-1/2}, 
\end{equation}%
and therefore we obtain%
\begin{eqnarray}
H &=&p_{b}\dot{b}-L \\
&=&\sqrt{\frac{2GM}{b}}\sqrt{1-\frac{\alpha ^{2}b}{2GM}}\left[ \frac{%
Gp_{b}^{2}}{2R\alpha ^{2}}-\frac{R\alpha ^{2}}{2G}\right] .  \notag
\end{eqnarray}%
Now, we calculate the Hamiltonian constraint in terms of $\dot{b}$
\begin{eqnarray}
H&=&\frac{R\alpha ^{2}}{2G}\sqrt{\frac{b}{2GM}}\left( 1-\frac{\alpha ^{2}b}{2GM%
}\right) ^{-1/2}\times\notag\\
&&\times\left[ \dot{b}^{2}-\frac{2GM}{b}\left( 1-\frac{\alpha ^{2}b}{%
2GM}\right) \right] =0,
\end{eqnarray}%
and immediately get the following solution
\begin{equation}
\dot{b}^{2}=\frac{2GM}{b}-\alpha ^{2},
\end{equation}%
which is exactly the equation (36). When the horizon radius, $r_{h}=\frac{%
2GM}{\alpha ^{2}}$, is much larger than the scale on which we are probing
the singularity, we can write
\begin{equation*}
1-\frac{\alpha ^{2}b}{2GM}\sim 1
\end{equation*}%
so the Hamiltonian takes the form
\begin{equation}
H=\sqrt{\frac{2GM}{b}}\left[ \frac{GP_{b}^{2}}{2R\alpha ^{2}}-\frac{R\alpha
^{2}}{2G}\right] .
\end{equation}%
The spatial volume
\begin{eqnarray}
V &=&\int drd\theta d\phi \sqrt{h}=4\pi Rab^{2}  \notag \\
&=&4\pi Rb^{3/2}\sqrt{2GM}\sqrt{1-\frac{\alpha ^{2}b}{2GM}}
\end{eqnarray}%
simplifies when using the above approximation and we obtain 
\begin{eqnarray}
V &=&l_{0}b^{3/2}  \notag \\
l_{0} &=&4\pi R\sqrt{2GM} \label{simplevolume}
\end{eqnarray}%
The canonical pair is given by $b\equiv x$ and $p_{b}$, with Poisson bracket $%
\left\{ x,p_{b}\right\} =1$.

For isotropic models, only holonomies evaluated in isotropic connections $%
A_{a}^{i}=\tilde{c}\delta _{a}^{i}$ appear. Along straight lines in the
direction of translation symmetries $X_{I}^{a}=\left( \partial /\partial
X^{I}\right) ^{a}$, the holonomies $\exp \left( \int X_{I}^{a}A_{a}^{i}\tau
_{i}\right) $ from the fundamental representation of $SU\left( 2\right) $ have
matrix elements of the form $\exp \left( i\mu c\right) $, where $\mu $
depends on the length of the curve used. Here, it turns out to be useful to
introduce $c:=V_{0}^{1/3}\tilde{c}$ defined in terms of the coordinate size $V_{0}$ of the region used to define the isotropic phase space \cite{21}.

Using this motivation we introduce the following function which will be used instead of the momentum (from now on we leave out the subscript $b$ for momentum associated with this observable) \cite{23}
\begin{equation}
U_{\gamma }\left( p\right) \equiv \exp \left( 8\pi G\frac{i\gamma }{L}%
p\right) 
\end{equation}%
where $\gamma $ is a real parameter and $L$ fixes the length scale. The
parameter $\gamma $ determines the separation of momentum points in the phase
space. 

The pair $\left( x,U_{\gamma }\left( p\right) \right) $ has the following Poisson bracket algebra%
\begin{equation}\label{Poisson-bracket}
\left\{ x,U_{\gamma }\left( p\right) \right\} =8\pi G\frac{i\gamma }{L}%
U_{\gamma }\left( p\right) 
\end{equation}%
A straightforward calculation gives 
\begin{eqnarray}
U_{\gamma }^{-1}\left\{ V^{n},U_{\gamma }\right\}  &=&l_{0}^{n}U_{\gamma
}^{-1}\left\{ \left\vert x\right\vert ^{3n/2},U_{\gamma }\right\}  \\
&=&i8\pi Gl_{0}^{n}\frac{\gamma }{L}\frac{3n}{2}sgn\left( x\right)
\left\vert x\right\vert ^{3n/2-1}  \notag
\end{eqnarray}%
We are concerned with the quantity $\frac{1}{\left\vert x\right\vert }$
which can serve as an indicator for singularity presence because classically it diverges for $\left\vert x\right\vert
\rightarrow 0$ thus producing singularity. From this moment we choose $n=1/3$ 
\begin{equation}\label{inverseroot}
\frac{sgn\left( x\right) }{\sqrt{\left\vert x\right\vert }}=-\frac{2Li}{8\pi
Gl_{0}^{1/3}\gamma }U_{\gamma }^{-1}\left\{ V^{1/3},U_{\gamma }\right\} .
\end{equation}

\subsection{Quantization}

We will use the basis of Hillbert space introduced in \cite{22,23}, which is
formed by eigenstates of \ $\hat{x}$. This implies the existence of a
self-adjoint operator $\hat{x}$, acting on the basis states according to 
\begin{equation}\label{x-operator}
\hat{x}\left\vert \mathbf{\mu }\right\rangle =L\mu \left\vert \mathbf{\mu }%
\right\rangle 
\end{equation}%
Next, we want to promote the classical momentum function $U_{\gamma }=e^{\left( 8\pi G\frac{i\gamma }{L}p\right) }$ to operator (note that it is not exactly holonomy of a connection but rather translation generator). We can do so by defining the action of $\hat{U}_{\gamma }$ on the basis states with the help of the
definition equation (\ref{x-operator}) and using commutation relation based on the Poisson bracket
between $x$\ and $U_{\gamma }$. Applying the natural definition of  $\hat{U}_{\gamma }$ as a translation operator \cite{24,21} and computing commutation relation we obtain
\begin{equation}
\hat{U}_{\gamma }\left\vert \mathbf{\mu }\right\rangle =\left\vert \mathbf{%
\mu -\gamma }\right\rangle ,\text{ \ \ \ \ \ \ \ \ \ \ \ \ \ }\left[ \hat{x},%
\hat{U}_{\gamma }\right] =-\gamma L\hat{U}_{\gamma }.
\end{equation}%
Using canonical quantization of Poison bracket $i\hbar \left\{ ,\right\} \rightarrow \left[ ,\right]$, and equation (\ref{Poisson-bracket}) we get a relation for the length scale
\begin{equation}
L=\sqrt{8\pi }l_{p}
\end{equation}

\subsection{Volume operator and disappearance of the singularity}

In the vicinity of the singularity we assume the approximate equation (\ref{simplevolume}). Then the volume operator acts in the following way on the basis states 
\begin{equation}
\hat{V}\left\vert \mathbf{\mu }\right\rangle =l_{0}\left\vert x\right\vert
^{3/2}\left\vert \mathbf{\mu }\right\rangle =l_{0}\left\vert L\mu
\right\vert ^{3/2}\left\vert \mathbf{\mu }\right\rangle 
\end{equation}%
Using the equation (\ref{inverseroot}) and promoting the Poisson
brackets to commutators, while setting $\gamma =1$, we find%
\begin{equation}\label{inv-operator}
\frac{\widehat{1}}{\left\vert x\right\vert }=\frac{1}{2\pi
l_{p}^{2}l_{0}^{2/3}}\left( \hat{U}_{\gamma }^{-1}\left[ \hat{V}^{1/3},\hat{U%
}_{\gamma }\right] \right) ^{2}.
\end{equation}%
On the basis states this operator acts in the following way
\begin{eqnarray}
\hat{U}_{\gamma }^{-1}\left[ \hat{V}^{1/3},\hat{U}_{\gamma }\right]
\left\vert \mathbf{\mu }\right\rangle  &=&\left( \hat{U}_{\gamma }^{-1}\hat{V%
}^{1/3}\hat{U}_{\gamma }-\hat{U}_{\gamma }^{-1}\hat{U}_{\gamma }\hat{V}%
^{1/3}\right) \left\vert \mathbf{\mu }\right\rangle   \notag \\
&=&l_{0}^{1/3}l_{p}^{1/2}\left( \sqrt{\mu -1}-\sqrt{\mu }\right) \left\vert 
\mathbf{\mu }\right\rangle 
\end{eqnarray}%
so finally we get 
\begin{equation}
\frac{\widehat{1}}{\left\vert x\right\vert }\left\vert \mathbf{\mu }%
\right\rangle =\sqrt{\frac{2}{\pi l_{p}^{2}}}\left( \sqrt{\mu -1}-\sqrt{\mu }%
\right) ^{2}\left\vert \mathbf{\mu }\right\rangle .
\end{equation}%
We can see that the spectrum is bounded from above and so the singularity is resolved in the quantum theory (the theory gives finite predictions for observables related to singularity). In fact, the eigenvalue of operator $\frac{\widehat{1}}{\left\vert x\right\vert }$ corresponding to the state $\left\vert \mathbf{0}\right\rangle 
$ which probes the classical singularity is equal to $%
\sqrt{\frac{2}{\pi l_{p}^{2}}}$, which is the highest eigenvalue of the spectrum.
Specifically, the operator corresponding to the curvature invariant 
\begin{eqnarray}
\mathcal{~R}_{\mu \nu \rho \sigma }\mathcal{R}^{\mu \nu \rho \sigma } &=&%
\frac{48M^{2}G^{2}}{r^{6}}+\frac{128M\pi G^{2}\eta ^{2}}{r^{5}}+\frac{%
256G^{2}\pi ^{2}\eta ^{4}}{r^{4}}  \notag \\
\equiv \frac{48M^{2}G^{2}}{b\left( t\right) ^{6}}&+&\frac{128M\pi G^{2}\eta
^{2}}{b\left( t\right) ^{5}}+\frac{256G^{2}\pi ^{2}\eta ^{4}}{b\left(
t\right) ^{4}}
\end{eqnarray}%
is then automatically finite in quantum mechanics. Promoting it to operator and evaluating on $%
\left\vert \mathbf{0}\right\rangle $\ we get 
\begin{eqnarray}\label{curvature}
&&\mathcal{~}\widehat{\mathcal{R}_{\mu \nu \rho \sigma }\mathcal{R}^{\mu \nu
\rho \sigma }}\left\vert \mathbf{0}\right\rangle \notag\\ 
&&=\left( \widehat{\frac{48M^{2}G^{2}}{\left\vert x\right\vert ^{6}}}+\widehat{\frac{128M\pi
G^{2}\eta ^{2}}{\left\vert x\right\vert ^{5}}}+\widehat{\frac{256\pi
^{2}G^{2}\eta ^{4}}{\left\vert x\right\vert ^{4}}}\right) \left\vert \mathbf{%
0}\right\rangle  \notag\\
&&=\left( \textstyle{\frac{384M^{2}G^{2}}{\pi ^{3}l_{p}^{6}}+\sqrt{\frac{2}{\pi ^{5}}}%
\frac{512M\pi G^{2}\eta ^{2}}{l_{p}^{5}}+\frac{1024\pi ^{2}G^{2}\eta ^{4}}{%
\pi ^{2}l_{p}^{4}}}\right) \left\vert \mathbf{0}\right\rangle 
\end{eqnarray}%
On the other hand, when $\left\vert \mu \right\vert \rightarrow \infty $ the
eigenvalue of $\frac{\widehat{1}}{\left\vert x\right\vert }$ goes to zero which is natural behaviour for large $\left\vert x\right\vert $.

One should note that the above result comes from expressing the operator $\frac{\widehat{1}}{\left\vert x\right\vert }$ via relation (\ref{inv-operator}). The resolution of singularity is not given by the existence of minimal length because the operator $\hat{x}$ contains zero in its (continuous) spectrum as can be seen from (\ref{x-operator}). The classical singularity is removed because $\frac{\widehat{1}}{\left\vert x\right\vert }$ has finite eigenvalue even when eigenvalue of $\hat{x}$ vanishes. This is in complete agreement with LQC. However, as pointed out in \cite{Thiemann}, the situation is more complicated when considering full Loop Quantum Gravity. 

One should also make sure the Kretschmann curvature scalar operator spectrum is finite at the classical singularity which is, however, implied by its simple form for our model. The result (\ref{curvature}) confirms both the resolution of singularity (in contrast to classical behavior) and vanishing of curvature in asymptotic region in quantum picture (in agreement with classical behavior, as one would expect far from singularity).   

Also, it is possible to show that the quantum Hamiltonian constraint gives a
discrete difference equation for the coefficients of the physical states.

\section{Conclusion}
We have seen that we have not been successful in removing the naked singularity by using relativistic quantum mechanics (for both Klein-Gordon and Dirac equations). On the other hand we have shown that the curvature singularity of general global monopole is resolved when the geometry is quantized using loop techniques. Unfortunately, one cannot directly compare the results because the loop quantization relied on radial coordinate being timelike beneath the horizon which is not the case for naked singularity of pure monopole. But still, this might be an indication that the first method is not reliable for determining the fate of singularities in quantum theory and one should rather focus on quantization of the geometry itself. But even the approach inspired by loop quantization that relied on restricted class of geometries should not be trusted completely. One should allow, e.g., for deviations from spherical symmetry to be completely sure about the fate of singularities.

\begin{acknowledgements}
T.T. gratefully acknowledges the hospitality of Institute of Theoretical Physics (Charles University in Prague) during her stay. O. S. was supported by grant GA\v{C}R 14-37086G.
\end{acknowledgements}

\appendix
\section{Geometric quantities}

The spatial metric is 
\begin{equation}
h_{ij}=\left( a^{2}\left( t\right) ,b^{2}\left( t\right) ,b^{2}\left(
t\right) \sin ^{2}\theta \right) ,
\end{equation}%
The extrinsic curvature is $K_{ij}=-\frac{1}{2}\frac{\partial h_{ij}}{%
\partial t}$, and so%
\begin{gather}
K_{ij}=-\left( a\dot{a},b\dot{b},b\dot{b}\sin ^{2}\theta \right) ,  \notag \\
K=K_{ij}h^{ij}=-\left( \frac{\dot{a}}{a}+2\frac{\dot{b}}{b}\right) \\
K_{ij}K^{ij}=\left( \frac{\dot{a}^{2}}{a^{2}}+2\frac{\dot{%
b}^{2}}{b^{2}}\right) \text{\ \ \ \ \ \ \ }  \notag \\
K_{ij}K^{ij}-K^{2}=-2\left( \frac{\dot{b}^{2}}{b^{2}}+2\frac{\dot{a}\dot{b}}{%
ab}\right)
\end{gather}%
The Ricci curvature for the space section is 
\begin{equation}
^{\left( 3\right) }R=\frac{2}{b^{2}}.
\end{equation}

\end{document}